\documentclass{article}

\usepackage{PRIMEarxiv}

\usepackage[utf8]{inputenc} 
\usepackage[T1]{fontenc}    
\usepackage{hyperref}       
\usepackage{url}            
\usepackage{booktabs}       
\usepackage{amsfonts}       
\usepackage{nicefrac}       
\usepackage{microtype}      
\usepackage{lipsum}
\usepackage{fancyhdr}       
\usepackage{graphicx}       

\graphicspath{{media/}}     

\title{Nonlinearities of Penman and Penman-Monteith Equations across Multiple Timescales
\thanks{\textit{\underline{Citation}}: 
\textbf{Authors. Title. Pages.... DOI:000000/11111.}} 
}

\author{
  Yizhi Han\\
  1\\
  School of Hydrology and Water Resources, Nanjing University of Information Science and Technology \\
  Nanjing, China\\
  \texttt{\{Author1, Author2\}email@email} \\
  \And
  Salvatore Calabrese \\
  2 \\
  Department of Biological and Agricultural Engineering, Texas A\&M University, College Station \\
  TX, USA\\
  \texttt{email@email} \\
  \And
  Huihua Du \\
  3 \\
  State Key Laboratory of Hydrology-Water Resources and Hydraulic Engineering, Nanjing Hydraulic Research Institute \\
  Nanjing, China\\
  \texttt{email@email} \\
  \And
  Jun Yin \\
  4 \\
  School of Hydrology and Water Resources, Nanjing University of Information Science and Technology \\
  Nanjing, China\\
  \texttt{email@email} \\
}

\begin{document}
\maketitle

\begin{abstract}
The nonlinear Penman and Penman-Monteith equations, widely used for estimating surface evapotranspiration at regional and global scales, were derived from turbulent transport of heat fluxes and thus apply to subhourly scale. However, these equations have been directly applied with hydrometeorological variables averaged at longer time intervals, leading to biases due to their nonlinearities. To address this problem, we used global eddy covariance flux data and Taylor expanded Penman and Penman-Monteith equations to explore their nonlinear components and the biases associated with the timescales mismatch. We found relatively small biases when applying Penman equation at longer timescale, in which the biases in equilibrium evapotranspiration mainly stem from the temperature-radiation covariance, whereas the biases in evapotranspiration due to drying power of air primarily come from the higher-order terms. Most of these biases can be corrected by linear regressions of first-order approximations. For Penman-Monteith equations, the corresponding biases are relatively larger but can be significantly reduced when daytime median stomatal conductance is used along with the first-order approximation of Penman-Monteith equation, suggesting the importance of diurnal variation of latent heat fluxes. The nonlinearity explored here serves as a reminder of the mismatched timescales for applying Penman and Penman-Monteith equations.
\end{abstract}

\keywords{First keyword \and Second keyword \and More}

\section{Introduction}

On average, global evapotranspiration is directly associated with approximately 82 W m$^{-2}$ of latent heat flux, equivalent to a quarter of incoming solar radiative flux at the top of the Earth's atmosphere \cite{wild2017global}. These latent heat fluxes are critical for the boundary layer dynamics, atmospheric convection, and cloud formation, playing an important role in controlling the Earth's energy balance and climate system. At regional scales, evapotranspiration is an important component of the water budget, thus influencing local water resources management and ecological conservation. Due to its inherent interlinkage with the latent heat flux, modeling evapotranspiration requires coupling surface energy balance to the equations of turbulent transport of heat fluxes. Penman \cite{penman1948natural} linearized the Clausius-Clapeyron relation to obtain the explicit expressions for evapotranspiration from wet surface. 
Monteith\cite{monteith1965evaporation} further extended this approach to find the evapotranspiration from non-wet surface with assumption of saturated condition inside the stomates. Penman and Penman-Monteith equations have been widely used in various fields ranging from regional water resources management to global climate projections.

\begin{figure}
    \centering
    \includegraphics[width=\textwidth]{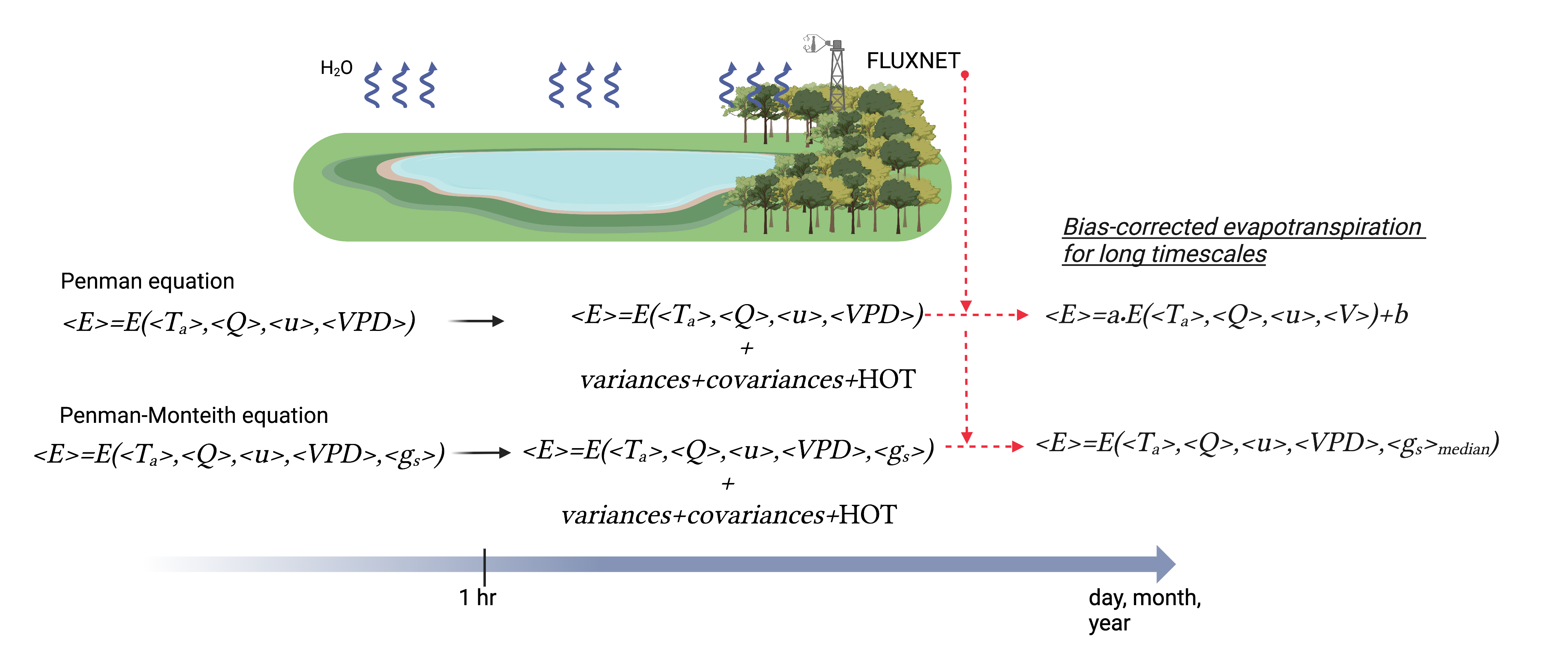}
    \caption{Biases and corrections of Penman and Penman-Monteith equations across timescales}
    \label{fig:intro}
\end{figure}

Penman and Penman-Monteith equations were derived from the turbulent transport of heat fluxes, whose time frame is about half an hour (see Fig. \ref{fig:intro}). Hence, directly applying these equations at longer timescales may result in biased results due to the nonlinearities in these equations. While this problem seems obvious, it has not received much attention and is often ignored. For example, most climate models in Coupled Model Intercomparison Projects provide outputs at monthly time step; yet these data have been widely used to calculate evapotranspiration or related variables without explicitly addressing the mismatch with the timescale of Penman and Penman-Monteith equations (hereafter the ``timescale mismatch"). It is in fact reported that potential evapotranspiration calculated from Penman equation by using data of different time steps can have large biases in certain locations. For example, Yang and Zhong \cite{yang2011} compared the differences in evapotranspiration calculated by using monthly and daily average data at three meteorological stations in China and found that the former was 2.0 \% higher;  Perera et63
al. \cite{perera2015} compared the evapotranspiration calculated by using hourly and daily data over the Australia and found the root mean square difference is around 0.26 mm day$^{-1}$; Itenfisu et al. \cite{itenfisu2003} suggested that potential evapotranspiration based on hourly meteorological data is 8\% higher than these from daily data for 49 sites across the United States; Suleiman and Hoogenboom \cite{suleiman2009} compared daily evapotranspiration from 15-minute and daily data in Georgia, USA, and found that the former was 8$\%$ higher. These case studies have been mostly restricted to data comparisons without systematically explaining the reasons for these biases. They thus offer limited information on the nonlinearities of the Penman or Penman-Monteith equations and how they affect their applications in hydrometeorological studies.

Towards this goal, here we Taylor expanded the Penman and Penman-Monteith equations to investigate their nonlinearities and explore how they can lead to biases in evapotranspiration predictions. Leveraging eddy covariance flux data of FLUXNET at high temporal resolutions over the world \cite{pastorello2020fluxnet2015}, we found that the biases when applying Penman equation are associated with temperature-radiation covariance and higher-order terms, most of which can be corrected by linear regression. The biases when applying Penman-Monteith equation are much larger but can be significantly reduced by using the daytime median stomatal conductance (see Fig. \ref{fig:intro}). The rest of the article is organized as follows. In section 2, we explain the timescale of Penman equation, which is then used to split evapotranspiration into the two parts related to radiation and drying power of air to explore their nonlinearities and the associated biases in sections 3 and 4. Following the same approach, we further investigate Penman-Monteith equation in section 5. Some practical method for correcting these biases are reported in section 6. Final conclusions are summarized in section 7.
\section{Penman and Penman-Monteith Equations}
The derivation of Penman equation starts from the turbulent transport of latent and sensible heat fluxes from surface to the atmosphere \cite{brutsaert2013evaporation, porporato2022ecohydrology}
\begin{equation}
H=\rho c_p \overline{w'T'}=\rho c_{p} g_h({T}_0-\overline{T}_a) \label{eq:HH}
\end{equation}\begin{equation}
\label{eq:EE}
\rho_{w} E=\rho\overline{w'q'}=\rho g_e\frac{\epsilon}{p_0}\left[e_0-\overline{e}_a\right]
\end{equation}
where $H$ is sensible heat flux, $E$ is evapotranspiration rate, $c_p$ is specific heat capacity, $\rho_w$ is water density, $\rho$ is air density,  $\epsilon$  ($\approx$ 0.622) is ratio of the water and air molar masses, $p_0$ is atmospheric pressure, $e$ is vapor pressure, $q\, (=\epsilon e/p_0)$ is specific humidity, $T_0$ is the surface temperature, the bars indicate Reynolds averages at timescale of half hour to an hour \cite{stull1988Introduction}, and subscripts 0 and $a$ refer the corresponding variables at surface and in atmosphere (e.g., low level of the mixing layer), $T_a$ and $e_a$ are temperature and vapor pressure in the atmosphere, and $g_h$ and $g_e$ are the conductances to water vapor and sensible heat transfer. For vegetated surface, $g_e$ typically consists of a series of stomatal and atmospheric conductances $g_ag_s/(g_a+g_s)$, whereas $g_h$ is essentially aerodynamic conductance $g_a$. Under near-neutral condition with approximately logarithmic wind profile, the aerodynamic conductance can be modeled as \cite{garratt1994atmospheric, brutsaert2013evaporation}
\begin{equation}
\label{eq:ga}
g_a=\frac{\bar{u}_a k^2}{\ln \left[\left(z_a-d\right) / \epsilon^*\right] \ln \left[\left(z_a-d\right) / \epsilon_q^*\right]}
\end{equation}
where $k=0.41$ is the von Karman constant, $\bar{u}$ is wind speed at $z_a$, $d$ is displacement height, $\epsilon^*$ and $\epsilon_q^*$ are the momentum and water vapor roughness heights.
For wet surface, we expect open stomates with high stomatal conductance ($g_s \gg g_a$) and saturation water vapor pressure inside the stomates following Clausius-Clapeyron relationship ($e_0=e_\mathrm{sat} (T_0)$). Combining Eqs. (\ref{eq:HH}) and (\ref{eq:EE}) with the energy balance equation ($Q=H+\rho_w\lambda_wE$, where $Q$ is available energy), one can numerically calculate the evapotranspiration rate for given meteorological measurements. Due to the presence of the nonlinear function $e_\mathrm{sat}(T)$ in Eq. (\ref{eq:EE}), Penman \cite{penman1948natural} approximated it as a linear function of temperature and found evapotranspiration rate as
\begin{equation}
\label{eq:pm}
\rho_{w} E=\underbrace{\frac{\Delta}{\lambda_{w}\left(\Delta+\gamma^{*}\right)} Q}_{\rho_{w} E_{e}}+\underbrace{\frac{\gamma^{*}}{\Delta+\gamma^{*}} E_{A}}_{\rho_wE_{d}},
\end{equation}
and
\begin{equation}
\label{eq:EA}
E_{A}=\frac{\epsilon}{p_{0}} \rho g_{a}[e_\mathrm{sat}(T_a)-e_a]=\frac{\epsilon}{p_{0}} \rho g_{a} V,
\end{equation}
where $\gamma^*$ is psychrometric constant, $\lambda_w$ is latent heat of water vaporization, $V=(e_\mathrm{sat}(T_a)-e_a)$ is vapor press deficit (VPD), $E_A$ is drying power of the air, and $\Delta=de_\mathrm{sat}/dT|_{T=T_a}$ is the slope of the saturation vapor pressure. Note that more accurate solution can be found by using Lambert-W function \cite{mccoll2020practical}. The first term is often referred to as equilibrium evapotranspiration $E_e$ and the second term is evapotranspiration due to drying power of air $E_d$. 

For non-wet surface, we expect lower water vapor pressure inside the stomates controlled by leaf water potential. However, Monteith further extended Penman's combination approach by assuming water vapor pressure inside the stomates is still saturated but using stomatal conductance, $g_s$, to model the reduced evapotranspiration rate. \cite{monteith1965evaporation}
\begin{equation}
\label{eq:pmon}
\rho_w E=\frac{\Delta}{\lambda_w\left[\Delta+\gamma^*\left(1+{g_a}/{g_s}\right)\right]} Q+\frac{\gamma^*}{\Delta+\gamma^*\left(1+{g_a}/{g_s}\right)} E_A,
\end{equation}
which reduces to the Penman equation (\ref{eq:pm}) for complete opening of stomata (e.g., $g_s \gg g_a$). Similar to Penman equation (\ref{eq:pm}), Penman-Monteith should also be applied at sub-hourly timescale;

While it is common to omit the bars of Reynolds average in Eq. (\ref{eq:HH}) and (\ref{eq:pmon}), Penman and Penman-Monteith equations still remain valid at a subhourly timescale. It should also be noted that linearizing saturation vapor pressure in Penman equation certainly introduces biases and readers may refer to the literature for more details (e.g.,Gao et al. \cite{gao1988applications}; Milly\cite{ milly1991refinement}; McColl\cite{mccoll2020practical}). Here we only focus on the timescale and nonlinearities of these equations.

Applying subhourly Penman and Penman-Monteith equations at longer timescales simply by using daily, monthly, or annual (DMA) averages of meteorological variables introduces biases at the corresponding timescale. As shown in Appenidx A, for any given nonlinear function $f: \mathbf{R}^n \rightarrow \mathbf{R}$, the average of the function is not the same as the function of the averages of the independent variables. The difference depends on the variances, covariances, and higher moments of the independent variables as analyzed in the next sections. 

\section{Temperature-Radiation Covariances Accounting for Biases in $E_e$}
\label{sec:eq}
We analyze the nonlinearity of the first part of Penman equation, $E_e$, which is a bivariate nonlinear function of $T_a$ and $Q$, i.e., $E_e=f(T_a, Q)$. Applying Taylor series in Appendix A to $f(T_a, Q)$ and acknowledging the second-order derivative with respect to $Q$ is zero (i.e., $f_{QQ}=0$) yields
\begin{equation}
\label{eq:TQmean}
\langle f(T_a,Q)\rangle=\underbrace{f(\langle T_a\rangle,\langle Q\rangle)}_{\mathrm{I}}+\underbrace{1/2f_{T_{a}T_{a}}\sigma_{T_a}^2}_{\mathrm{II}}+\underbrace{f_{T_{a}Q}\mathrm{cov}(T_a,Q)}_{\mathrm{III}}+\underbrace{\langle \mathrm{H.O.T.}\rangle}_{\mathrm{IV}},
\end{equation}
where $f(T_a,Q)$ refers to $E_e$ at subhourly timescale, and $\langle \cdot \rangle$ is different from Reynolds averages and refers to the averages for given periods. Therefore, $\langle f(T_a,Q)\rangle$ is the average of $f(T_a,Q)$ at longer timescales (e.g., DMA), which can be regarded as the exact value. Term I $f(\langle T_a\rangle,\langle Q\rangle)$, as the approximation, is the function of the averages of $T_a$ and $Q$ at the corresponding timescales. According to Eq. (\ref{eq:TQmean}), the errors between the two depend on the temperature variances (II), temperature-radiation covariances (III), and higher-order terms (IV).

To quantify the importance of each term in Eq. (\ref{eq:TQmean}), we substituted sub-hourly $T_a$ and $Q$ from FLUXNET2015 dataset into Eq. (\ref{eq:pm}) to calculate sub-hourly $E_e$ and then averaged at DMA scales. Since surfaces of near these observation sites are not always wet, the results calculated from Penman equations only refer to the potential evapotranspiration as the atmospheric water demand  \cite{allen1998crop, zomer2022version}. The sub-hourly FLUXNET2015 data were also used to calculate the variance and covariance (i.e, $\sigma^2_{T_a}$ and cov($T_a, Q$)) at DMA scales. Missing values on any variables were excluded in the following analysis. Finally, these values were used to calculate each terms in Eq. (\ref{eq:TQmean}) to assess the nonlinearity of the equilibrium part of the Penman equation. Note that term IV is calculated as the residual of the whole equation, i.e., Term IV = $\langle f(T_a,Q)\rangle$ $-$ Term I $-$ Term II $-$ Term III. 

\begin{figure}
\begin{center}
\includegraphics[width=\textwidth]{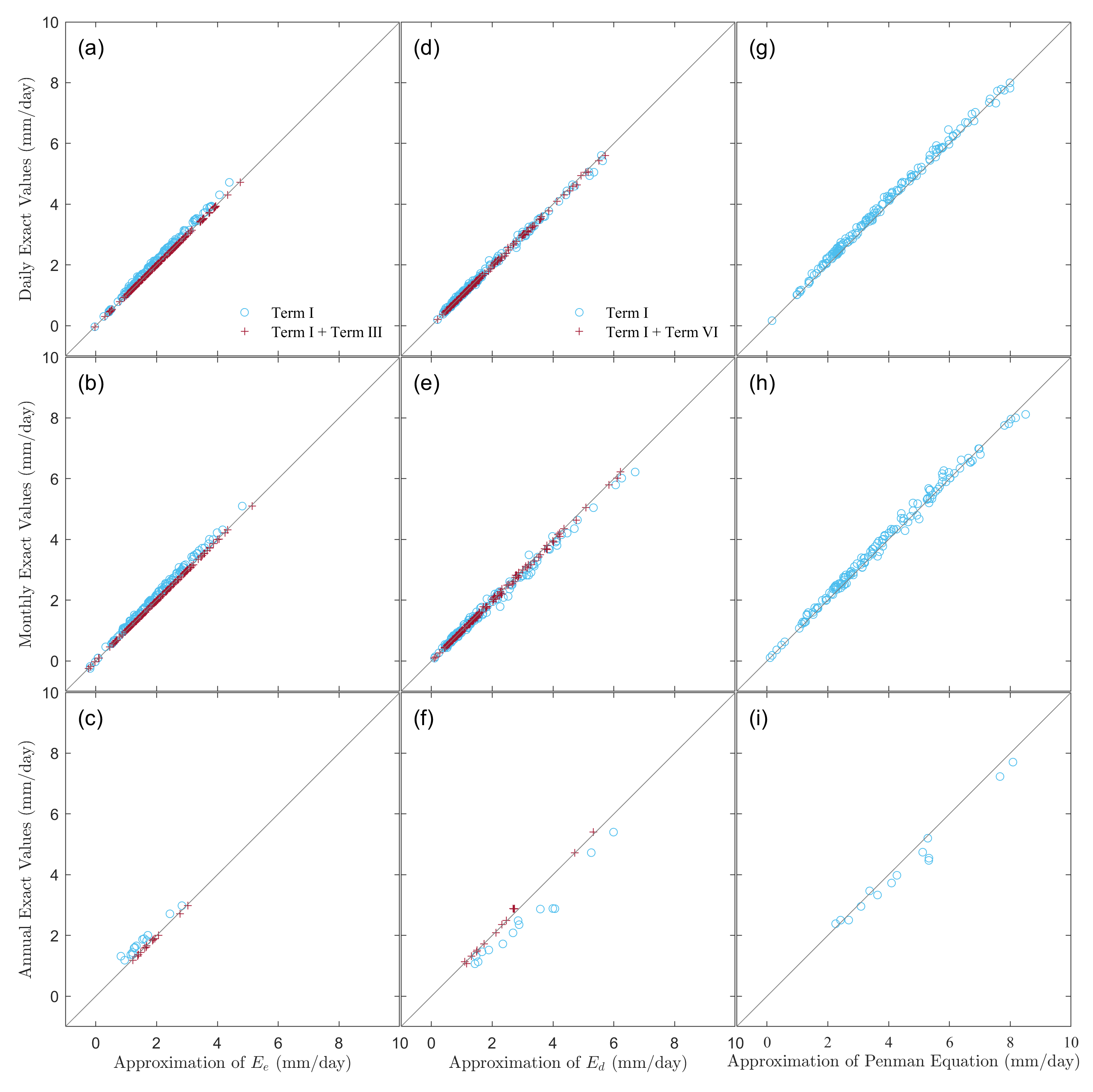}
\end{center}
\caption{Approximation of Penman equations at varying timescales. In panels (a)-(c), the exact values of equilibrium evapotranspiration $E_e$ are compared with first-order approximations (i.e., Term I, blue circles) along with temperature-radiation covariance terms (i.e., Term I + Term III, red crosses) at the (a) daily, (b) monthly, and (c) annual scales, respectively. In panels (d)-(f), the exact values of $E_d$ are compared with the first-order approximations (i.e., Term I, blue circles) along with the higher-order terms (i.e., Term I +Term VI, red crosses) at the (d) daily, (e) monthly, and (f) annual scales, respectively. In panels (g)-(i), the exact values of $E_e+E_d$ are compared with the first-order approximations (i.e., sum of Term I in Eq. (\ref{eq:TQmean}) and Term I in Eq. (\ref{eq:TuVPDmean})).}
\label{fig:approx}
\end{figure}

As shown in Fig. \ref{fig:approx} a-c, the approximation (i.e, term I $f(\langle T_a\rangle,\langle Q\rangle)$) at DMA timescales are quite close to the exact value (i.e., $\langle f(T_a,Q)\rangle$), although they become less accurate at longer timescales (RMSE are 0.15, 0.16, 0.29 mm/day at DMA scales respectively). It is also clear that the biases increase roughly linearly with the approximations with non-zero intercepts at annual timescale. To quantify the importance of each term, we combined term I with other terms and compared them with the exact values. As shown in Fig. \ref{fig:approx} a-c, the addition of term III makes the approximation very close to the 1:1 line with root-mean square errors (RMSE) of 0.01, 0.02, 0.04 mm/day at DMA scales respectively), suggesting the covariance term accounts for a significant portion of the biases.

This dominant covariance term seems to be always positive, which can be explained by the positive signs of both the derivative and the covariance. Solar radiation warms the Earth surface and increases the air temperature, leading to the positive correlation between radiation and temperature, $\mathrm{cov}(T_a,Q)>0$. The derivative, expressed as
\begin{equation}
    f_{T_a,Q}=\frac{d}{dT_a}\frac{\Delta}{\lambda_w(\Delta+\gamma^*)},
\end{equation}
is also positive because of the monotonically increasing functions of $\Delta/(\Delta+\gamma^*)$ and $\Delta(T_a)$.
The positive covariance term therefore results in underestimation for term I at different timescales (i.e., above the 1:1 line in Fig. \ref{fig:approx} a-c).

\section{Higher-Order Terms Associated with Most Biases in $E_d$}
\label{sec:vpd}
The second part of Penman equation, $E_{d}=g(T_a,u,V)$, is a nonlinear function of three variables. Similar to Sec. \ref{sec:eq}, we find the Taylor-series expansion of $g(T_a,u,V)$, see Appendix A, and consider that the second-order derivatives with respect to $u$ and $V$ are zeros (i.e., $g_{uu}=0$ and $g_{V,V}=0$). The resulting equation is
\begin{eqnarray}
\label{eq:TuVPDmean}
&\langle g(T_a,u,V)\rangle = \underbrace{g(\langle T_a\rangle,\langle u\rangle,\langle V\rangle)}_{\mathrm{I}}+\underbrace{1/2g_{T_{a}T_{a}}\sigma_{T_a}^2}_{\mathrm{II}}+\underbrace{g_{T_{a}u}\mathrm{cov}(T_a,u)}_{\mathrm{III}}\nonumber\\ &+\underbrace{g_{T_{a}V}\mathrm{cov}(T_a,V)}_{\mathrm{IV}}+\underbrace{g_{uV}\mathrm{cov}(u,V)}_{\mathrm{V}}+\underbrace{\langle \mathrm{H.O.T.}\rangle}_{\mathrm{VI}},
\end{eqnarray}
where $\langle g(T_a,u,V)\rangle$ is the average of $g(T_a,u,V)$ at longer timescale and refers to the true values calculated from Penman equation. Term I, $g(\langle T_a\rangle,\langle u\rangle,\langle V\rangle)$, is the approximation at longer timescales. The differences between true values and its approximation are therefore associated with temperature variances (II), temperature-wind speed covariances (III), temperature-VPD covariances (IV),  wind speed-VPD covariances (V), and higher-order terms (VI). A detailed analysis of each item is given below.

Following the same approach as in Sec. \ref{sec:eq}, we calculate each term in Eq. (\ref{eq:TuVPDmean}) using the FLUXNET2015 dataset (see Fig. \ref{fig:approx} d-f). The approximations (term I, $g(\langle T_a\rangle,\langle u\rangle,\langle V\rangle)$) at DMA scales are also close to the exact values ($\langle g(T_a,u,V)\rangle$) and tend to be larger at longer timescales. The biases also increase roughly linearly with the approximation with non-zero intercepts at annual timescale. Unlike the equilibrium evapotranspiration, the H.O.T. seems to be the most important component of the biases, which change from slightly positive to negative with increasing timescales (i.e., term I is slightly above 1:1 line at daily timescale and below 1:1 line at annual timescale).

\section{Non-additive Effects of $g_s$ in Penman-Monteith Equation}
To estimate the evapotranspiration from non-wet surface, we may use Penman-Monteith equation, which is also expected to be applied at the subhourly timescale. However, it has been extended to longer timescales (e.g., Orgaz et al. \cite{orgaz2007model}; Whitley et al. \cite{whitley2009comparing}). The biases due to the timescale mismatch were investigated in this section following the same approach in Secs. \ref{sec:eq} and \ref{sec:vpd}.

While it seems that evapotranspiration is explicitly expressed as function of hydrometeorological variables, the calculation of $g_s$, for example from Jarvis equation \cite{jarvis1976interpretation}, involves leaf water potential, which is seldom measured and required to be coupled to a soil-plant-atmosphere continuum model. It can be treated as explicit only when $g_s$ is empirically modeled as other measurable variables, such as soil moisture. Here, we calculate bulk stomatal conductance by inversion of Eq. (\ref{eq:pmon}) \cite{grace1995fluxes}
\begin{equation}
    g_s=\frac{\gamma^*g_a\rho_wE}{\Delta Q/\lambda_w+\gamma^*E_A-\Delta\rho_wE-\gamma^*\rho_wE}.
\end{equation}
This allows us to focus on Penman-Monteith equation itself in this study without directly modeling $g_s$, which is outside the scope of the work here.

\begin{figure}
    \centering
    \includegraphics[width=\textwidth]{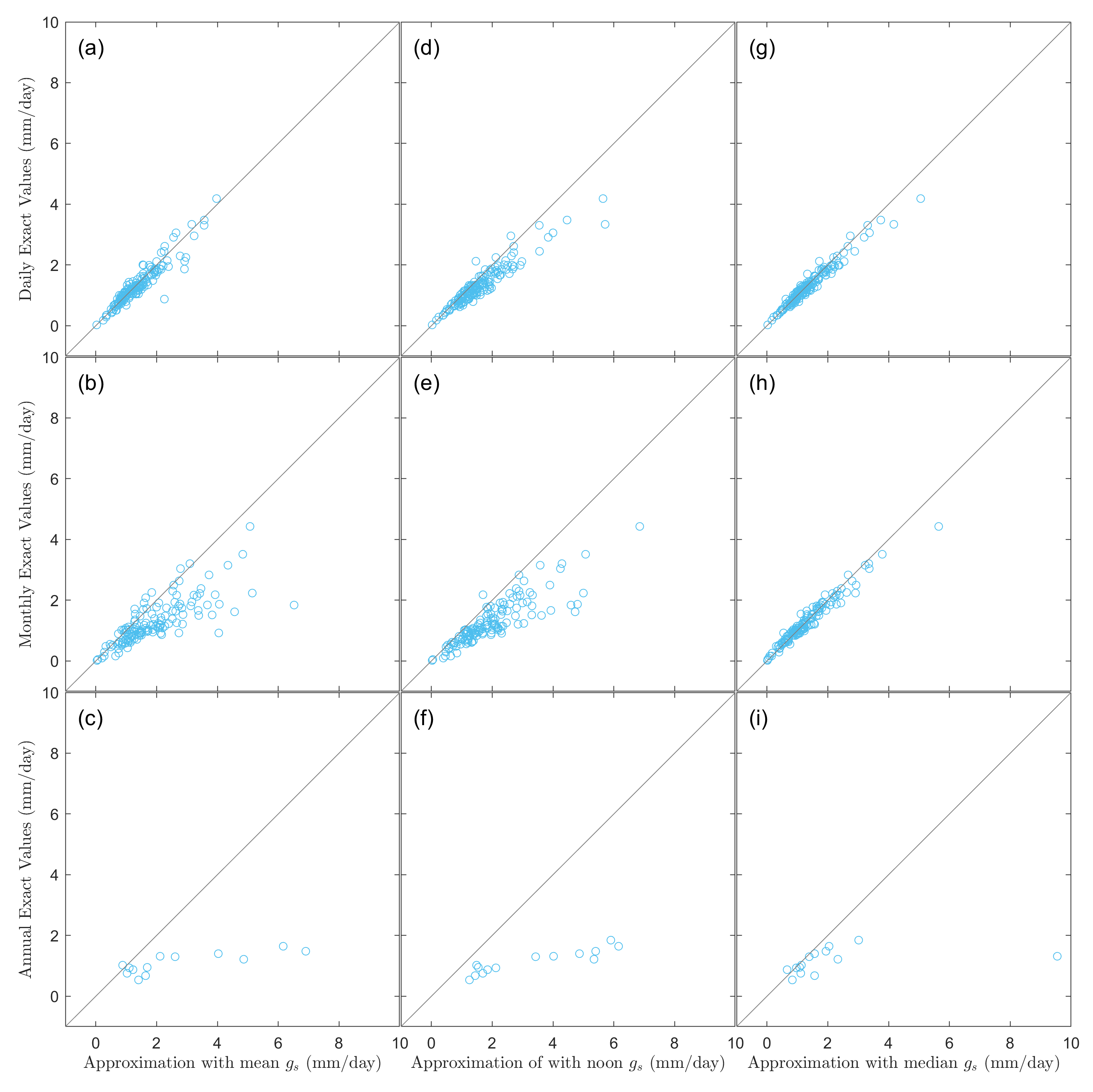}
    \caption{Approximation of Penman-Monteith equations at varying timescales. In panels (a)-(c), the exact values of evapotranspiration are compared with first-order approximations with mean values $g_s$ at the (a) daily, (b) monthly, and (c) annual scales, respectively. Similarly, first-order approximations with midday $g_s$ and daytime median $g_s$ are presented in panels (d)-(f) and (g)-(i), respectively.}
    \label{fig:p-m}
\end{figure}

Since there are five variables, its Taylor series to the second order can be as many as 16 terms, complicating the decomposition of the biases. In this regard, we only compared the exact evapotranspiration (i.e., averages over DMA scales) and the first-order approximations (i.e., the first term of Taylor series with averages of $g_s$ and other hydrometeorological variables at DMA scales). For $g_a$, it is estimated from Eq. (\ref{eq:ga}) with observed canopy heights for FLUXNET data \cite{wang2020determinants}. The results from FLUXNET data are reported in Fig. \ref{fig:p-m} a-c. As can be seen, the biases are generally much larger than those in Penman equation, highlighting the strong nonlinearity of the Penman-Monteith equation.

\begin{figure}
    \centering
    \includegraphics[width=\textwidth]{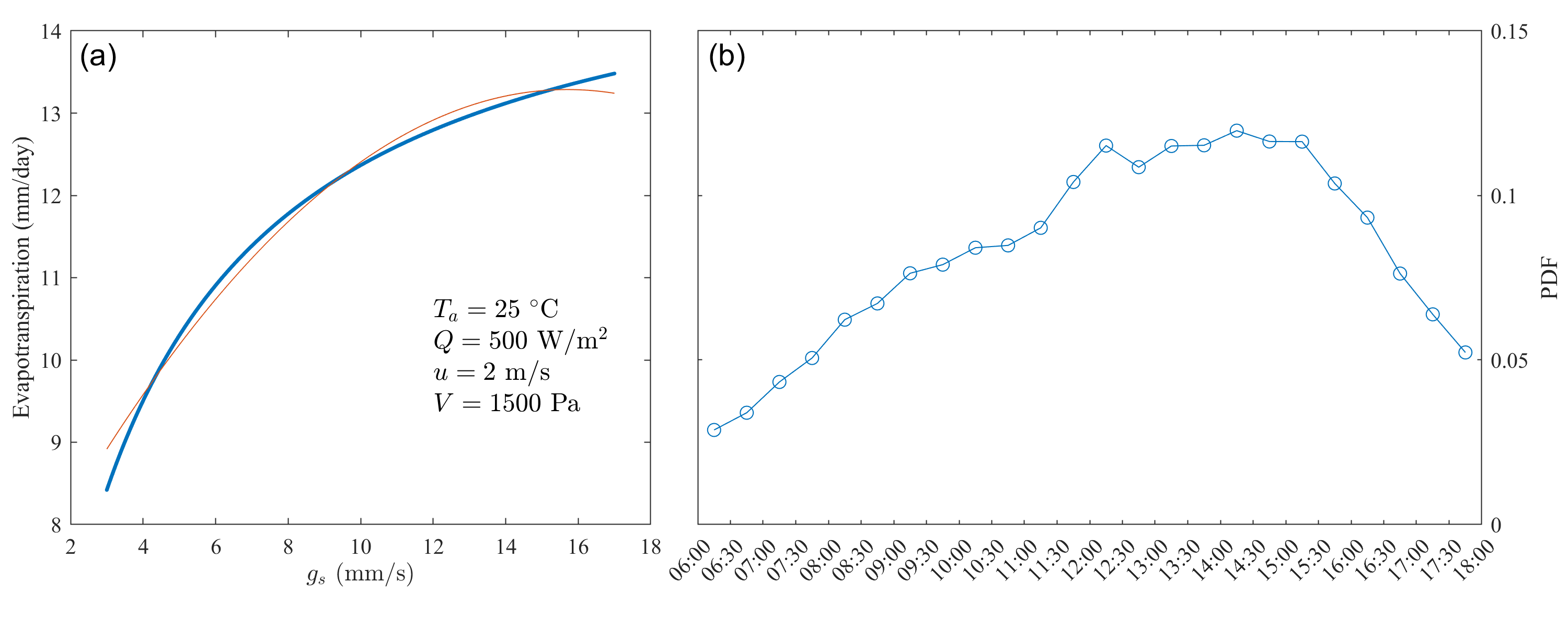}
    \caption{(a) Dependence of evapotranspiration on $g_s$ from Penman-Monteith equation under typical hydrometeorological conditions as indicated in the panel. (b) Probability density function(PDF) of the timing of daytime median $g_s$.}
    \label{fig:gs}
\end{figure}

To test if this nonlinearity of $g_s$ is additive, we set all hydrometeorological variables as typical values and analyzed the evapotranspiration merely as a function of $g_s$ (see Fig. \ref{fig:gs} a). This nonlinear function with convex shape can be well approximated by a quadratic function at certain range of $g_s$. If we further Taylor expand Penman-Monteith equation to the second order only with respect to $g_s$, the resulting approximations are close to those of first order approximations, suggesting that the nonlinearity is not additive but associated with the combining effects $g_s$ and all other variables.

\section{Correcting Evapotranspiration across Timescales}
\label{sec:corrections}
The overall potential evapotranspiration can be found by adding Eqs. (\ref{eq:TQmean}) and (\ref{eq:TuVPDmean}). As shown in Fig. \ref{fig:approx} g-i, the first terms in both equations can be used as the approximation for the accurate values. As analyzed above, linear regression could be used to correct most of the biases
\begin{equation}
\label{eq:lrap}
    \langle f(T_a,Q)\rangle+\langle g(T_a,u,V)\rangle
=a f(\langle T_a\rangle,\langle Q\rangle)+b g(\langle T_a\rangle,\langle u\rangle,\langle V\rangle) + c,
\end{equation}
where $a=1.05$, $b=0.96$, $c=0.11$ at daily scale, $a=1.06$, $b=0.93$, $c=0.10$ at monthly scale, and $a=1.21$, $b=0.81$, $c=-0.04$ at annual scale. One can also combine the first two terms together and have a simplified correction formula as
\begin{equation}
    \langle f(T_a,Q)\rangle+\langle g(T_a,u,V)\rangle
=a \left[ f(\langle T_a\rangle,\langle Q\rangle)+ g(\langle T_a\rangle,\langle u\rangle,\langle V\rangle)\right] + b,
\end{equation}
where $a= 0.99$, $b=0.16$ at daily scale, $a= 0.98$, $b=0.16$ at monthly scale,$a= 0.90$, $b=0.15$ at annual scale. This correction is slightly less accurate than Eq. (\ref{eq:lrap}) but has one less parameter.

For Penman-Monteith equation, its strong nonlinearity leads to larger biases in evapotranspiration for timescale mismatch. It might be inappropriate to use the average $g_s$ to represent the typical value during the period of consideration. In fact, diurnal variations of stomatal conductance or Bowen ratio simulated by atmospheric boundary layer models are nearly constant after the morning transit and before the sunset \cite{de_bruin_model_1983, porporato2009atmospheric, rigby2015approximate}. To check if typical daytime $g_s$ can be used for approximation in Penman-Monteith equation, we further compared the exact values with the approximations using midday $g_s$ and daytime (6 am - 6 pm) median $g_s$. The results, reported in Fig. \ref{fig:p-m} d-f and g-i, show that the biases were slightly reduced for midday $g_s$ but can be significantly reduced for median $g_s$ with RMSE reduced to 0.18, 0.19 and 2.27 mm day$^{-1}$ at DMA scales. The median $g_s$ is typically around noon and early afternoon (see Fig. \ref{fig:gs} b), when the heat fluxes and air temperature approach the maximum. Typical $g_s$ such as its median value explored here can be used to simplify the modeling of diurnal cycle of atmospheric boundary layer \cite{rigby2015approximate} and assist the analysis of land-atmosphere interaction.

\section{Conclusions}
In this study, we explored the nonlinearities of Penman and Penman-Monteith equations to address the problems associated with timescale mismatches. We Taylor expanded Penman and Penman-Monteith equations to compare its first-order approximations with the corresponding variance, co-variance, and higher-order terms. Using eddy covariance measurements from FLUXNET2015 datasets, we found that the overall biases for evaluating potential evapotranspiration using first-order approximations of Penman equation are relatively low and tend to be larger at longer timescales. The biases for equilibrium evapotranspiration mainly come from the temperature-radiation covariance term, whereas the biases for evapotranspiration due to the drying power of air are primarily associated with higher-order terms. Importantly, these biases can be reduced by linearly regressing potential evapotranspiration against the first-order terms of Penman equation with specific coefficients for each different timescale.

Larger biases were observed in the Penman-Monteith equation for the calculation of evapotranspiration from non-wet surface, as the addition of stomatal conductance significantly increases the nonlinearity of the Penman-Monteith equation for the first-order approximations. We showed that such biases can be considerably reduced when using the median daytime stomatal conductance, thus highlighting the large portion of latent heat flux during the daytime for typical diurnal variation of atmospheric boundary layer. From a practical point of view, the result of this work explain the nonlinear component of Penman and Penman-Monteith equations and provide some practical adjustments to reduce biases for their application at various timescales.

Finally, it should be kept in mind that this study focused on the Penman and Penman-Monteith equations and did not address biases associated with the assumptions in these equations such as lineariztion of saturation vapor pressure and saturated conditions inside the stomates of non-wet surfaces. The modeled stomatal conductance also has certain biases associated with the accuracy of inverting Penman-Monteith equation. Exploration of these assumptions and use of more comprehensive datasets at high spatial and temporal resolutions and with global coverage may be an interesting future extension of this work. 

\section*{Appendix A: Taylor Series for multi-variable functions}
Application of subhourly Penman equation at longer timescale should proceed with caution. To explain this point, we consider the general case of two nonlinear function, $h(x,y)$ and $h(x,y,z)$. Taylor expanding the first function at the mean values of $x$ and $y$ , and the second one at the mean values of $x$ , $y$ and $z$  gives
\begin{eqnarray}
\label{eq:xy}
h(x,y)&=h(\langle x\rangle,\langle y\rangle)+(x-\langle x\rangle)h_x+(y-\langle y\rangle)h_y\nonumber\\ 
&+1/2(x-\langle x\rangle)^2h_{xx}+1/2(y-\langle y\rangle)^2h_{yy}\nonumber \\
&+(x-\langle x\rangle)(y-\langle y\rangle)h_{xy} +\mathrm{H.O.T.},
\end{eqnarray}
and
\begin{eqnarray}
\label{eq:xyz}
h(x,y,z)&=h(\langle x\rangle,\langle y\rangle,\langle z\rangle)+(x-\langle x\rangle)h_x+(y-\langle y\rangle)h_y\nonumber+(z-\langle z\rangle)h_z\nonumber \\
&+1/2(x-\langle x\rangle)^2h_{xx}+1/2(y-\langle y\rangle)^2h_{yy}\nonumber+1/2(z-\langle z\rangle)^2h_{zz}\nonumber \\
&+(x-\langle x\rangle)(y-\langle y\rangle)h_{xy}+(x-\langle x\rangle)(z-\langle z\rangle)h_{xz}\nonumber \\
&+(y-\langle y\rangle)(z-\langle z\rangle)h_{yz} +\mathrm{H.O.T.}, 
\end{eqnarray}
where $\langle x\rangle$, $\langle y\rangle$ and $\langle z\rangle$, different from Reynolds averages, refer to the averages of $x, y, z$ for any given period, function $h$ with subscripts are the corresponding derivatives and evaluated at  $(\langle x\rangle,\langle y\rangle, \langle z\rangle)$, and H.O.T. are higher-order Taylor series terms. Applying averages for the whole equations (\ref{eq:xy} and \ref{eq:xyz}) yields
\begin{equation}
\label{eq:xymean}
\langle h(x,y) \rangle = h(\langle x\rangle,\langle y\rangle)+1/2h_{xx}\sigma_x^2+1/2h_{yy}\sigma_y^2+h_{xy}\mathrm{cov}(x,y) +\langle \mathrm{H.O.T.} \rangle,
\end{equation}
and 
\begin{eqnarray}
\label{eq:xyzmean}
&\langle h(x,y,z) \rangle =h(\langle x\rangle,\langle y\rangle,\langle z\rangle)
+1/2h_{xx}\sigma_x^2+1/2h_{yy}\sigma_y^2+1/2h_{zz}\sigma_z^2+h_{xy}\mathrm{cov}(x,y)\nonumber\\ 
&+h_{xz}\mathrm{cov}(x,z) +h_{yz}\mathrm{cov}(y,z) +\langle \mathrm{H.O.T.} \rangle,
\end{eqnarray}
where $\sigma_x$, $\sigma_y$ and $\sigma_z$ are standard deviations of $x$, $y$ and $z$,  $\mathrm{cov}(x,y)$ is the covariance of $x$ and $y$, similarly, $\mathrm{cov}(x,z)$ and $\mathrm{cov}(y,z)$ are the covariance of variables corresponding to them respectively. In linear system, e.g., $h(x,y)=x+y$, Eq. (\ref{eq:xymean}) is reduced to $\langle h(x,y) \rangle =h(\langle x\rangle,\langle y\rangle)$, and for $h(x,y,z)=x+y+z$, Eq. (\ref{eq:xyzmean}) is reduced to $\langle h(x,y,z) \rangle =h(\langle x\rangle,\langle y\rangle,\langle z\rangle)$; In multiplicative system, e.g., $h(x,y)=xy$, Eq. (\ref{eq:xymean}) is reduced to $\langle h(x,y) \rangle =h(\langle x\rangle,\langle y\rangle) +h_{xy}\mathrm{cov}(x,y)$, and for $h(x,y,z)=xyz$, Eq. (\ref{eq:xyzmean}) is reduced to $\langle h(x,y,z) \rangle =h(\langle x\rangle,\langle y\rangle) +h_{xy}\mathrm{cov}(x,y) +h_{xz}\mathrm{cov}(x,z) +h_{yz}\mathrm{cov}(y,z)$; for Penman equation, most terms in Eq. (\ref{eq:xymean}) and Eq. (\ref{eq:xyzmean}) reserve.

\section*{Data Availability}
FLUXNET data used in study are available from \url{https://fluxnet.org/data/fluxnet2015-dataset/}.

\section*{Author Contributions}
Conceptualization: S.C., and J.Y.

Data curation: Y.H.

Formal Analysis: Y.H., S.C., H.D., and J.Y.

Funding acquisition: J.Y

Investigation: Y.H., S.C., H.D., and J.Y.

Methodology: Y.H., S.C., H.D., and J.Y.

Project administration: J.Y.

Resources: J.Y.

Supervision: J.Y.

Visualization: Y.H.

Writing – original draft: Y.H. and J.Y.

Writing – review \& editing: Y.H., S.C., H.D., and J.Y.

\section*{Acknowledgments}
J.Y. acknowledges support from the Natural Science Foundation of Jiangsu Province (BK20221343) and NUIST startup funding (1441052001003).
\bibliographystyle{unsrt}  
\bibliography{references}

\end{document}